# Near-infrared negative-index metamaterials consisting of multiple layers of perforated metal-dielectric stacks

Shuang Zhang,[1] Wenjun Fan,[1] N. C. Panoiu,[2] K. J. Malloy,[1] R. M. Osgood,[2] and S. R. J. Brueck[1]*

[1]*Center for High Technology Materials and Department of Electrical and Computer Engineering, University of New Mexico, Albuquerque, New Mexico 87106, USA*
[2]*Department of Applied Physics and Applied Mathematics, Columbia University, New York, New York 10027, USA*

**Abstract: In this paper, we numerically demonstrate a near-infrared negative-index metamaterial (NIM) slab consisting of multiple layers of perforated metal-dielectric stacks and exhibiting low imaginary part of index over the wavelength of negative refraction. The effective index is obtained using two different numerical methods and found to be consistent. Backward phase propagation is verified by calculation of fields inside the metamaterial. These results point to a new design of low loss thick metamaterial at optical frequencies.**

More than 40 years ago, Veselago proposed many unconventional phenomena for a medium with negative refractive index, as well as the application of a NIM slab as an imaging lens.[1] Recently, Pendry took a step further and claimed that the NIM slab can magnify the evanescent field and work as a perfect lens. [2] This, as well as the first experimental demonstration of an artificial NIM in the microwave region, [3] led to numerous efforts in this research topic.

In the last two years, there has been much progress towards the realization of negative-index metamaterial (NIM) at optical frequencies. This includes the numerical studies on infrared magnetic metamaterials and NIMs [4, 5, 6] and experimental demonstration of THz, mid-infrared and near-infrared magnetic metamaterials [7, 8, 9, 10, 11] as well as near-infrared NIMs. [12, 13] The experimentally demonstrated near-IR NIMs in Ref. [12, 13] exhibit very large imaginary part of index, which make them not very suitable for real applications. To reduce the loss and enhance the transmission, more optimized structures similar to that in Ref [12] were numerically studied and experimentally verified. [14, 15] As shown in Ref. [14], a thin metamaterial ($\ll\lambda$) slab consisting of metal/dielectric/metal multiple layers with periodic array of holes exhibits negative refractive index at near-infrared wavelengths. This thin metamaterial slab is considered as a basic artificial molecule of negative-index metamateiral and can be used as a building block to construct a much thicker metamaterial ($\sim\lambda$), so that more interesting phenomena can be studied. Furthermore, the question as to how the optical properties of a metamaterial consisting of many unit cells differ from that of a single unit cell is frequently asked, a numerical study of the metamaterial with unit cells ranging from one to many will not only answer this question, but also give us more insight into the design of negative index metamaterials at optical frequencies.

As shown in Ref. [14, 16], an array of metallic stripe pairs along the direction of magnetic field vertically separated by a dielectric layer exhibits strong magnetic resonance with negative effective permeability over a certain wavelength range. The resonance wavelength has a linear relation with the stripe linewidth. Adding an array of metallic stripes along the direction of electric field to this magnetic structure resulting in metal/dielectric/metal films with 2D array of square holes perforating the multiple layers. Reduced surface plasma frequency in perforated metal films has been proposed and experimentally verified. [17, 18] Thus, by a careful design, a moderately negative permittivity can be achieved, which, in combination with the magnetic response, leads to negative refractive index. The schematic of the thick metamaterial structure simulated in this paper is shown in Fig. 1 with all the geometric parameters shown in the figure.

Rigorous coupled wave analysis (RCWA) was used for the simulation.[19] To extract the refractive index of this structure, two different methods are carried out. The first one is the determination of the effective index with the complex coefficients of transmission and reflectance on a metamaterial slab. In this method, the inverse of a cosine function is involved, which causes ambiguity due to the multiple branches of it. To resolve the ambiguity, metamaterials with different number of unit cells are simulated; the branch that is consistent for different numbers of unit cells is chosen as the root, as described in Ref. [20]

For light propagating in a periodic structure, infinite numbers of modes exist for a given frequency (each mode is the linear combination of many plane waves based on the periodicity in the transverse directions), while most of them decay very quickly and can be ignored. As the number of unit cell along the wave propagation direction gets larger, only the one with the smallest imaginary part dominates. [21] Based on this consideration, the second method is to solve the eigenvalues of the transfer matrix for a single unit cell, the mode with the smallest decay dominates and effective index associated with this mode is obtained.

In RCWA, the electromagnetic fields are expanded into a spatial Fourier series based on the period of the structures in the transverse directions. Suppose inside the periodic structures the forward and backward propagating light have spatial Fourier coefficients $A_{mn}$, $B_{mn}$, where $m$ and $n$ are the diffraction orders in $x$ and $y$ directions, the electrical field associated with the forward and backward propagation can be summed over all the spatial components. For instance, the electric field of the forward propagating beams can be expressed as:

$$\vec{E}_A = \sum_{m,n} A_{mn,\sigma} f_{mn}(x,y) g_{imn}(z) \hat{e}_{imn,\sigma} \tag{1}$$

with

$$f_{mn}(x,y) = e^{j\alpha_m x + j\beta_n y} \tag{2}$$

$$g_{imn}(x,y) = e^{j\Lambda_{imn} z} \tag{3}$$

where $\Lambda_{imn} = \sqrt{k^2 - \alpha_m^2 - \beta_n^2}$. Here the same number of diffraction orders *(2N+1)* are kept in both directions [a total number of *(2N+1)²*], with *m* and *n* ranging from (-*N*) to *N*. (*N* is limited by the computing power, here *N*=9). σ represents the polarization, TE or TM.

By using RCWA, the transmission matrix can be obtained numerically for a periodic structure, which relates the coefficient vector $A_{mn}$, $B_{mn}$ (before the unit cell) and $A'_{mn}$, $B'_{mn}$ (after the unit cell) by,

$$\begin{bmatrix} A_{mn} \\ B_{mn} \end{bmatrix} = M \begin{bmatrix} A'_{mn} \\ B'_{mn} \end{bmatrix} \tag{4}$$

Where *M* is a transfer matrix of dimension *4(2N+1)²*, which is calculated based on the Fourier transform of the dielectric function along the transverse directions. Next, the eigenvalues of the matrix **M** can be solved as $\beta_q$, where *q* is from 1 to *4(2N+1)²*, only half of the eigenvalues with modulus larger than 1 represent the physical propagating modes, with each mode being the linear combination of all the spatial harmonics. When light propagates through many unit cells, only the fundamental one with the smallest |β| (smallest loss) dominates, e.g. the mode with |β| closest to 1. Thus, the effective index can be expressed as:

$$n' = -\frac{\angle \beta}{k_0 d} \qquad (5)$$

$$n'' = \frac{\log(|\beta|)}{k_0 d} \qquad (6)$$

In equation (5) and (6), $d$ is the thickness of a single unit cell along the propagation.

We first numerically calculate the transmission and reflectance for one, two, five, six and ten layers of unit cells with both the incident and exit media as air. The transmission and reflectance spectra are shown in Fig. 3. For a single layer, the transmission shows a dip around 2 µm and a peak around 1.93 µm. With increasing number of layers, the transmission at long wavelength ($\lambda > 1.93$ µm) decrease rapidly and approaching zero. For multiple layers of unit cells, over the wavelength below 1.95 µm, the T and R oscillate with the wavelength, characteristic of a pass band with finite thickness, which is the negative index region, as will be shown later. With the dimension of unit cell along the propagation being 130 nm, the thickness of the slab with ten unit cells is only 1.3 µm, which is less than the wavelength of interest. However, as many as eight resonant peaks are observed over the pass band from 1.5 µm to 1.95 µm. As will be shown later, the large number of resonances is due to the rapid increase of absolute value of effective index with wavelength over the negative refraction region.

Next, the effective indices of structures with different thicknesses are calculated using the complex coefficients of transmission and reflectance, as shown in Fig. 4. For a single unit cell, the real part of index is continuous and negative from 1.77 to 2.18 µm, the imaginary part shows a peak around 2 µm. For multiple numbers of unit cells, the negative refraction starts from 1.5 µm, the indices decrease quickly with wavelength to ~-8 at 2 µm. Over this negative index range, the real part of index for two to ten layers agrees well and the imaginary part of index for five to ten layers converges nicely and is very small over the range from 1.5 to 1.7 μm (less than 0.1 for ten layer unit cells). As mentioned before, the inverse method involves finding root for a cosine function, which has infinite branches. As shown in Fig. 4, real part of effective index along branch 1 is consistent for different numbers of unit cell up to 2 μm, when wavelength is longer than 2 μm, the branch 1 starts to diverge for different layers. However, we can find another branch, e.g. branch 2 as shown in Fig. 4, which is consistent for different numbers of unit cells. Although there is a large discontinuity between branch 1 and branch 2 at 2 μm, however, to meet the requirement that the refractive index has to be consistent for different number of unit cells of metamaterial, we need to accept these two branches at different wavelength regions. For wavelength over 2 μm (branch 2 is chosen), the real part of index is almost zero and the imaginary part is much larger than the real part, exhibiting metallic properties. Compared to single unit cell, the imaginary part is low and flat over this long wavelength range.

Discontinuity of refractive index with respect to wavelength does not occur in natural materials. To better understand this discontinuity, the second method, a modal analysis was carried out. By using equation (5) and (6), both the real and imaginary parts of the refractive index for the two modes with the smallest decay are obtained, as shown in Fig. 5 (a). The reason for the discontinuity is clear: the imaginary part of the first mode exceeds that of the second mode above 2 μm, for a thick slab, the second mode become the dominant mode. The results obtained from modal analysis are consistent with that from inverse method very well. In addition, the coupling efficiency of an incident beam into each mode is also important. As shown in Fig. 4, the effective index calculated for five to ten layers converge very well except for a narrow wavelength range around 2 μm where the two modes are comparable, indicating that the fundamental mode is absolutely dominant over other modes in most of the frequency range for such a thick slab. Furthermore, in Fig. 3, over the negative index region, the transmission is very large even for ten unit cells, indicating good coupling between a normally incident beam with the negative index fundamental mode. Next, we plot the figure of merit defined in Ref. [14], the ratio of real part to the imaginary part of the effective index, which is shown in Fig. 5 (b). The highest value of 25 is achieved around 1.7 µm. Compared to that of a single unit cell [14], the optical properties of NIM slab consisting of multiple layer are much more improved.

It is of great interest to study how light propagates inside the metamaterials. We simulated a structure consisting of three unit cells along the propagating direction and calculated the distribution of electric and magnetic fields at several positions from *z=0* to *z=130* nm (the thickness of one unit cell) indicated by Fig. 6 (a). The phase of electric field averaged over one transverse unit cell versus propagating distance for three different wavelengths are shown in Fig. 6 (b). For all the three wavelengths, the phase decreases along the propagation, directly demonstrating the property of back phase propagation. Furthermore, the absolute amount of phase change from *z=0* to *z=130* nm increases with increase wavelength, which is consistent with the effective index shown in Fig. 5. The magnitude and phase of electric field at *z=30* nm were plotted in Fig. 6 (c) and 6 (d). The electric field is mainly confined inside the rectangular aperture; the phase of electric field inside the aperture is more uniform than the magnitude. In both plots, the modulation at high spatial frequency is due to the finite diffraction orders that are kept in the RCWA simulation.

Finally, we plot the distribution of magnetic field, both magnitude and phase, across one transverse unit cell at *z=30* nm for three wavelengths at 1.5, 1.7 and 2.1 µm. For the shortest wavelength λ = 1.5 µm, which is far away from the metamaterial resonance around 1.9-2.0 µm, the magnitude of magnetic field is slightly higher in the area beneath the metal pattern than that in the air aperture, while the phase of the magnetic field beneath the metal lines along the magnetic field $H_y$ (defined as region II) is exactly $\pi$ shifted from that in the air aperture (defined as region I) and in the area beneath the thin metal wires along the electrical field $E_x$ (defined as region III), as shown in Fig. 7 (a) and (b). Although the magnetic field in region III is as strong as that in region II, however, the area is much smaller, leading to an overall opposing magnetic field to that in region I. This directly verifies the existence of magnetic activity at this wavelength, although not very strong. Furthermore, this result confirms one point made in Ref. [14], e.g. the increase of metal line width along the electric field would weaken the magnetic resonant strength. At longer wavelength of 1.7 μm that is closer to the resonance frequency, the magnetic field in region II and III gets much larger than that in region I, leading to a stronger magnetic activity and more negative index than that at 1.5 μm, as shown in Fig. 7 (c) and (d).

At a longer wavelength of 2.1 μm, the fundamental mode switches to mode 2. The corresponding magnetic field distribution also shows very different features from that of 1.5 and 1.7 μm. The magnetic field in region II has the same phase as that in region I, while the magnetic field in region III opposes to that in region I, but much weaker. So the overall magnetic field in II and III is positive with respect to that in region I. In addition, the magnetic field at 2.1 μm is much smaller than that at 1.5 and 1.7 μm, mainly because of the large imaginary part of refractive index at 2.1 μm.

In conclusion, we have numerically demonstrated a low loss negative-index metamaterial with a thickness of the same order as the free space wavelength (for ten layers) in the near-infrared region. Numerical study on the electromagnetic fields inside the metamaterial slab verifies backward phase propagation and strong magnetic activity at negative index region. Further studies (not shown here) show that if the thin air gaps are eliminated, the results would not be affected much. This structure can be fabricated with current lithography and process techniques.

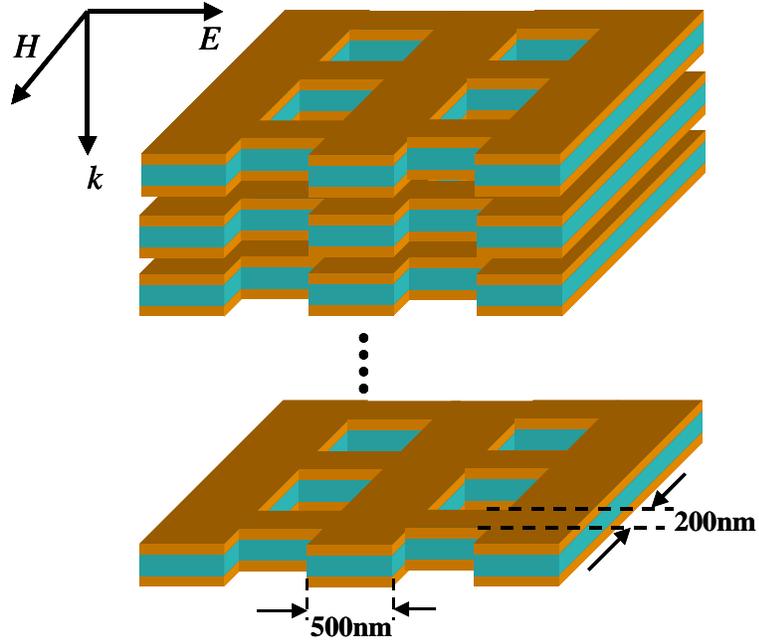

Fig. 1: Schematic of metamaterial consisting of multiple unit cells. the geometric parameters are: the pitch along in-plane directions is 801 nm, the linewidth of metal gratings along the direction of magnetic field is 500 nm and that along the electrical field is 200 nm. The thickness of the basic air/Au/dielectric/Au/air unit cell is 5/30/60/30/5 nm.

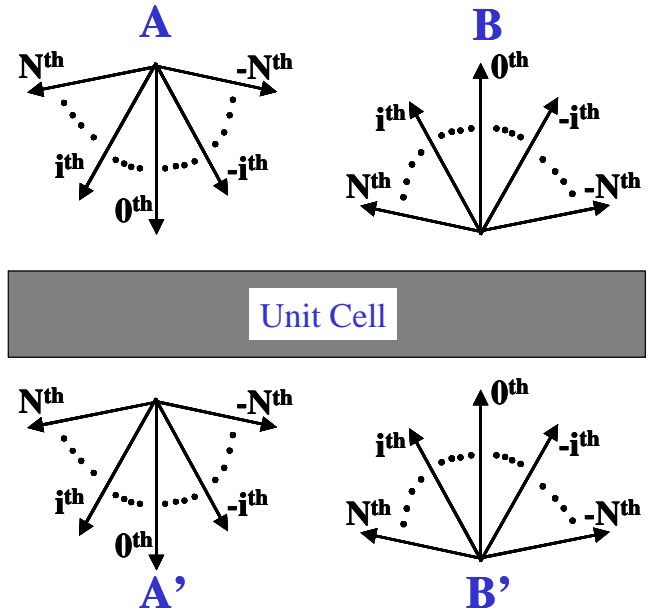

Fig. 2: Schematic of forward and backward propagating beams before and after one unit cell. (2N+1) diffraction orders are kept for both in-plane directions

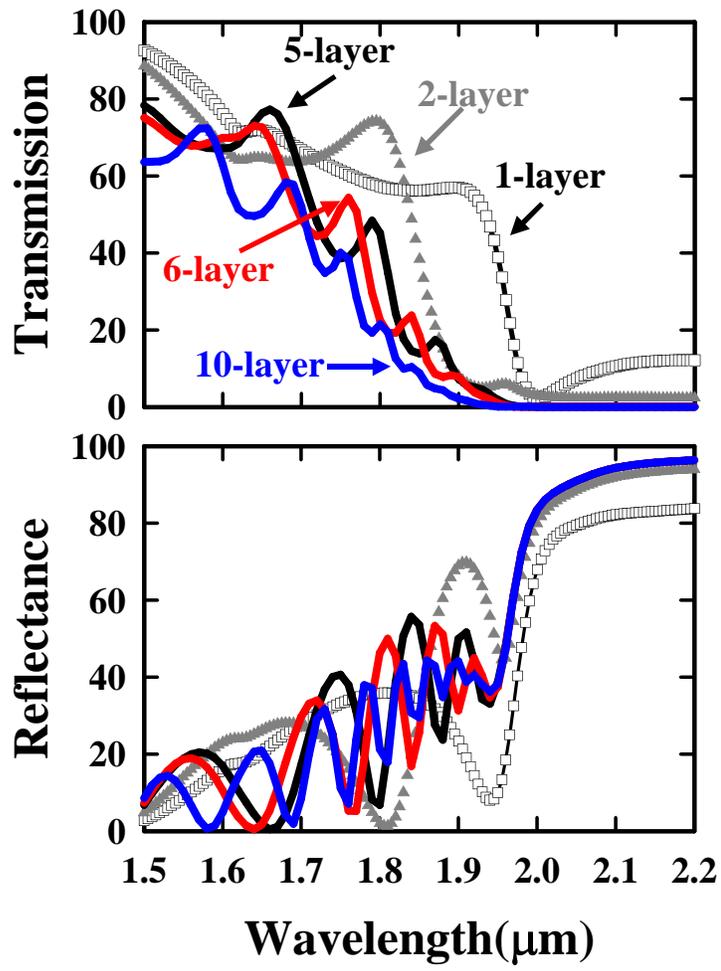

Fig. 3: the transmission (a) and reflectance (b) for 1, 2, 5, 6 and 10 layers of unit cells.

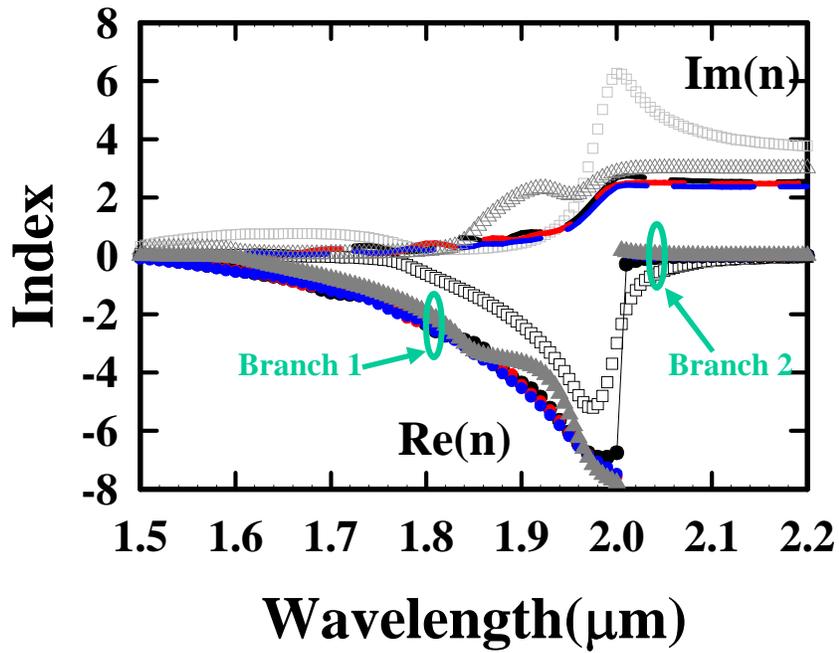

Fig. 4: Refractive index extracted from the complex coefficient of transmission and reflectance for different numbers of unit cells. The unfilled square, triangle, black, red and blue symbols represent the effective index extracted from 1, 2, 5, 6 and 10 layers respectively.

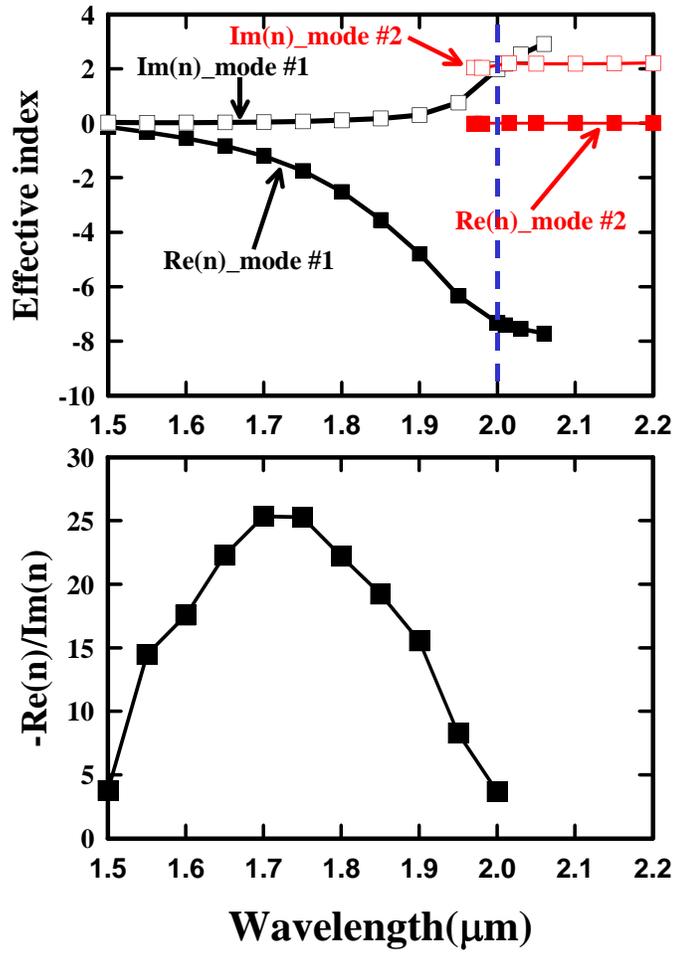

Fig. 5: (a) the effective index of two modes with the lowest decay. Mode 1 switch to mode 2 at the blue line when the imaginary part of it exceeds that of mode 2. (b) the ratio of the real part to the imaginary part of effective index.

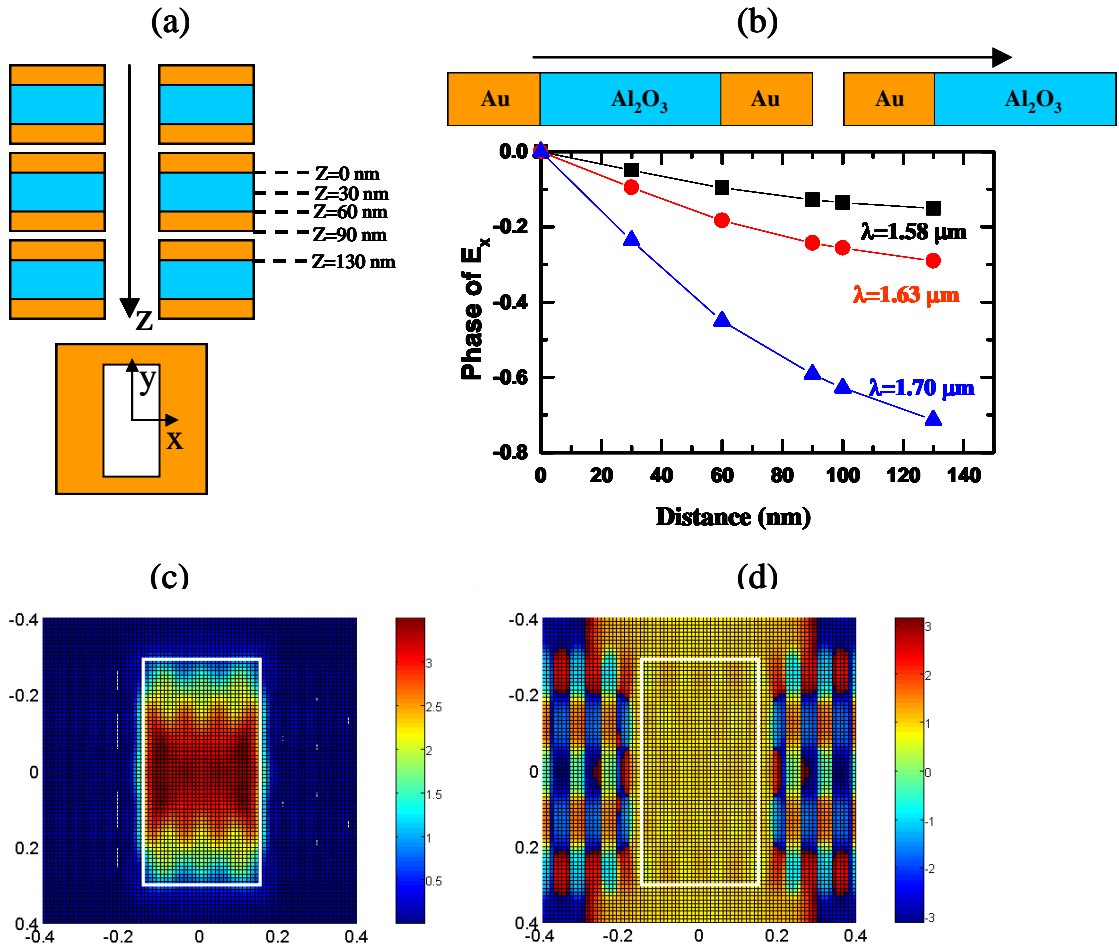

Fig. 6 (a): Schematic of light going through three unit cells along the direction of propagation. (b): Phase of average electric field at several positions from z=0 to z=130 as marked in (a). (c), (d): Field plot of electric field and magnetic field over one unit cell in transverse plane. White frames represent the size of the air aperture.

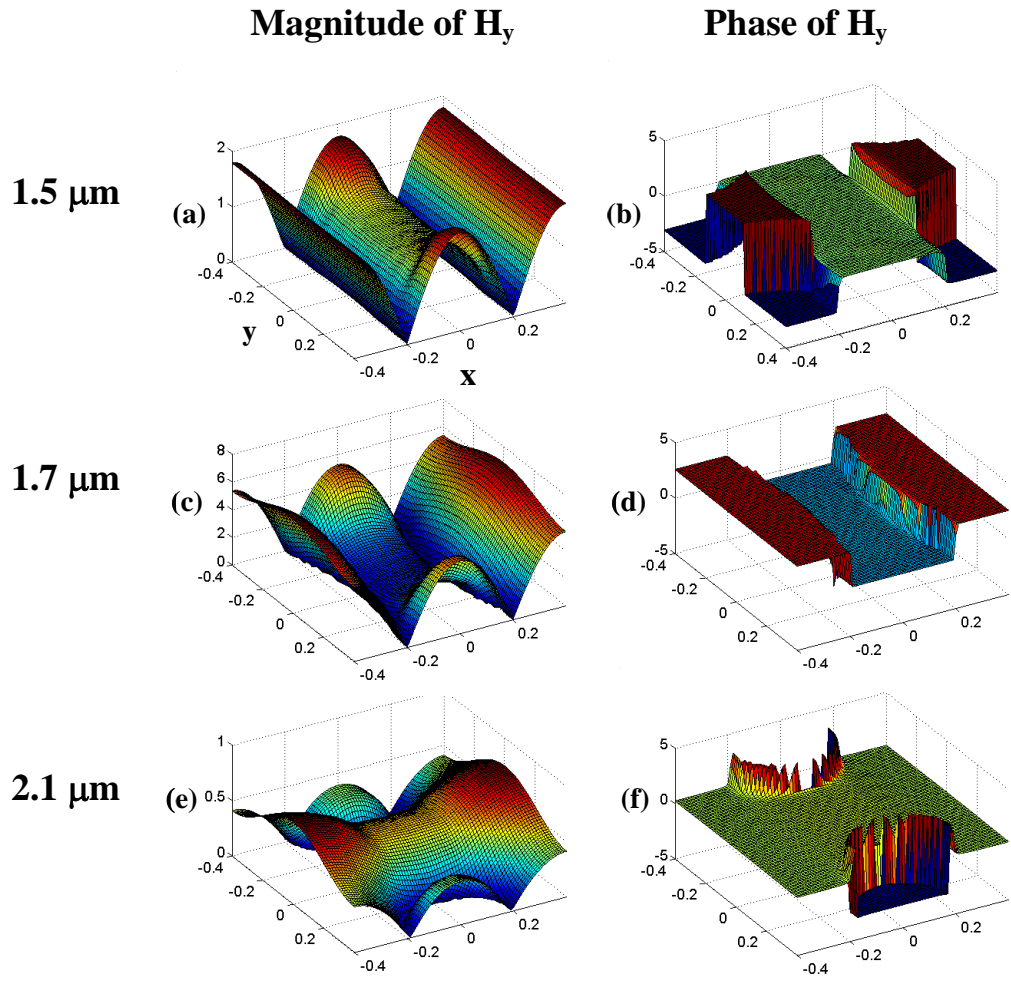

Fig. 7: The plot of magnitude and phase of magnetic field across one transverse unit cell at $z=30$ nm for wavelength at 1.5, 1.7 and 2.1 μm.

consisting of many unit cells